\documentclass[aps,prl,superscriptaddress,amsmath,%
                twocolumn,showpacs,10pt]{revtex4-1}
\usepackage{amssymb}
\usepackage{graphicx}
\usepackage{dcolumn}

\newcommand{\eq}[1]{Eq.\:(\ref{#1})}

\begin{document}
\title{High-Precision $f_{B_s}$
 and HQET from Relativistic Lattice QCD}
\author{C. McNeile}
\thanks{Current address: Theoretische Physik, Fachbereich C, Bergische 
Universit\'{a}t Wuppertal D-42097 Wuppertal, Germany 
(mcneile@uni-wuppertal.de)}
\affiliation{SUPA, School of Physics and Astronomy, University of 
Glasgow, Glasgow, G12 8QQ, UK}
\author{C. T. H. Davies}
\email[]{c.davies@physics.gla.ac.uk}
\affiliation{SUPA, School of Physics and Astronomy, University of 
Glasgow, Glasgow, G12 8QQ, UK}
\author{E. Follana}
\affiliation{Departamento de F\'{\i}sica Te\'{o}rica, Universidad de 
Zaragoza, E-50009 Zaragoza, Spain}
\author{K.Hornbostel}
\affiliation{Southern Methodist University, Dallas, Texas 75275, USA}
\author{G. P. Lepage}
\email[]{g.p.lepage@cornell.edu}
\affiliation{Laboratory of Elementary-Particle Physics, Cornell 
University, Ithaca, New York 14853, USA}
\collaboration{HPQCD collaboration}
\homepage{http://www.physics.gla.ac.uk/HPQCD}
\noaffiliation
\date{12 September 2011}
\pacs{11.15.Ha,12.38.Aw,12.38.Gc}

\begin{abstract}
We present a new determination of the $B_s$ leptonic decay constant 
from lattice QCD simulations that use gluon configurations from MILC 
and a highly improved discretization of the relativistic quark action 
for both valence quarks. Our result, $f_{B_s}\!=\!0.225(4)$\,GeV, is 
almost three times more accurate than previous determinations. We 
analyze the dependence of the decay constant on the heavy quark's mass 
and obtain the first empirical evidence for the leading 
$1/\sqrt{m_h}$~dependence predicted by Heavy Quark Effective Theory 
(HQET). As a check, we use our analysis technique to calculate the 
$m_{B_s}-m_{\eta_b}/2$~mass difference. Our result agrees with 
experiment to within errors of~$11\,\mathrm{MeV}$ (better than~2\%). We 
discuss how to extend our analysis to other quantities in $B_s$~and 
$B$~physics, making 2\%-precision possible for the first time.
\end{abstract}

\maketitle

Lattice simulations of QCD have become essential for high-precision 
experimental studies of $B$-meson decays\,---\,studies that test our 
understanding and the limitations of the standard model of weak, 
electromagnetic and strong interactions, and also determine fundamental 
parameters, like the CKM matrix, in that model. Accurate theoretical 
calculations of QCD contributions to meson masses, decay constants, 
mixing amplitudes, and semileptonic form factors are critical for this 
program, and lattice simulation is the main tool for providing these 
calculations. A major complication for the lattice simulations has been 
the large mass of the $b$~quark, which has necessitated the use of 
non-relativistic effective field theories like NRQCD to describe 
$b$~dynamics in the simulations. The need for effective field theories 
has made it difficult to achieve better than 5--10\%~precision for many 
important quantities.

Recently we overcame the analogous problem for $c$~quarks by 
introducing a highly improved discretization of the relativistic quark 
action that gives accurate results even on quite coarse lattices: the 
Highly Improved Staggered-Quark (HISQ) 
discretization~\cite{Follana:2006rc}. With this formalism, $c$~quarks 
are analyzed in the same way as $u$, $d$, and $s$~quarks, which greatly 
reduces the uncertainties in QCD~simulations of 
$D$~physics~\cite{Follana:2007uv,Davies:2009ih,Gregory:2009hq,McNeile:2010ji,Davies:2010ip,Na:2010uf}. More recently we showed that the HISQ action can be pushed to much higher masses\,---\,indeed, very close to the $b$~mass\,---\, using new lattices, from the MILC collaboration, with the smallest lattice spacing available today~($a\!=\!0.044$\,fm). This allowed us to extract a value for the $b$'s $\overline{\mathrm{MS}}$~mass that was accurate to better than~1\%. Here we extend that work in a new analysis of the $B_s$ meson's leptonic decay constant~$f_{Bs}$, which produces the most accurate theoretical value to date.

We also compute the mass difference~$m_{B_s}-m_{\eta_b}/2$, as an 
additional test of our analysis method. This difference is particularly 
sensitive to QCD dynamics because the leading (and uninteresting) 
dependence on the heavy quark's mass mostly cancels in the difference.

It would be quite expensive to extend our new analysis directly to 
$B$-meson quantities, because of the added costs associated with very 
light valence and sea quarks. This is unnecessary, however, because heavy-quark 
effective field theories like NRQCD already give accurate results for 
ratios of $B_s$~to $B$~quantities, like $f_{B_s}/f_B$. This is because 
the largest systematic errors from these effective theories, due to 
operator matching, cancel in such ratios. The ratio of the decay 
constants, for example, is known to~$\pm2$\% from effective field 
theories~\cite{bib-ratios}.
So the combination of accurate $B_s$ quantities from HISQ simulations, 
as discussed in this paper, with $B_s/B$ ratios from NRQCD or other 
effective field theories provides a potent new approach to 
high-precision $b$~physics generally. Note that no operator matching is 
required in our relativistic analysis of $f_{B_s}$ because of the exact 
chiral symmetry of the HISQ~formalism in the massless limit.

\begin{table}
   \caption{Parameter sets used to generate the 3-flavor gluon 
configurations
   analyzed in this paper. The lattice spacing is specified in terms
   of the static-quark potential parameter
    $r_1\!=\!0.3133(23)$\,fm\,\cite{Davies:2009tsa}; values
   for $r_1/a$ are from~\cite{MILC}.
   The bare quark masses are for the ASQTAD formalism and $u_0$
   is the fourth root of the plaquette. The spatial ($L$) and temporal
   ($T$) lengths of the lattices are also listed, as are the number
   of gluon configurations ($N_\mathrm{cf}$) and the number of time 
sources
   ($N_\mathrm{ts}$) per configuration used in each case.}
   \label{tab-cfg}
   \begin{ruledtabular}
      \begin{tabular}{cccccccc}
      Set & $r_1/a$ & $au_0m_{0u/d}$ & $au_0m_{0s}$ & $u_0$ & $L/a$ & 
$T/a$
      & $N_\mathrm{cf}\times N_\mathrm{ts}$
      \\ \hline
      1 & 2.152(5) & 0.0097 & 0.0484 & 0.860 & 16 & 48 & $631\times2$ 
\\
      2 & 2.618(3) & 0.01 & 0.05 & 0.868 & 20 & 64 & $595\times2$ \\
       3 & 3.699(3) & 0.0062 & 0.031 & 0.878 & 28 & 96 & $566\times4$\\
      4& 5.296(7) & 0.0036 & 0.018 & 0.888 & 48 & 144 & $201\times2$ \\
      5 & 7.115(20) & 0.0028 & 0.014 & 0.895 & 64 & 192  & 
$208\times2$\\
      \end{tabular}
   \end{ruledtabular}
\end{table}

\begin{table}
   \caption{Simulation results for each of the five configuration sets
    (Table~\ref{tab-cfg}) and
   several values of the heavy-quark's mass~$m_h$. The $s$-quark's
   mass~$m_s$ is tuned to be close to its physical value.
   Results are given for: the leptonic
   decay constant~$f_{H_s}$ and
    mass~$m_{H_s}$ of the pseudoscalar $h\overline{s}$~meson,
    and masses of the pseudoscalar $h\overline{h}$ and $s\overline{s}$
    mesons, $m_{\eta_h}$ and $m_{\eta_s}$ respectively.
    }
   \label{tab-data}
\begin{ruledtabular} \begin{tabular}{lllllll}%
 &  $am_s$ & $aM_{\eta_s}$ & $am_h$ & $aM_{H_s}$
& $af_{H_s}$ & $am_{\eta_h}$ \\
\hline
1 & 0.061 & 0.5049(4) & 0.66 & 1.3108(6) & 0.1913(7) & 1.9202(2) \\
 & 0.061 & 0.5049(4) & 0.81 & 1.4665(8) & 0.197(1) & 2.1938(2) \\
\hline
2 & 0.0492 & 0.4144(2) & 0.44 & 0.9850(4) & 0.1500(5) & 1.4240(1) \\
 & 0.0492 & 0.4144(2) & 0.63 & 1.2007(5) & 0.1559(7) & 1.8085(1) \\
 & 0.0492 & 0.4144(2) & 0.85 & 1.4289(8) & 0.161(1) & 2.2193(1) \\
\hline
3 & 0.0337 & 0.2941(1) & 0.3 & 0.7085(2) & 0.1054(2) & 1.03141(8) \\
 & 0.0337 & 0.2941(1) & 0.413 & 0.8472(2) & 0.1084(2) & 1.28057(7) \\
 & 0.0337 & 0.2941(1) & 0.7 & 1.1660(4) & 0.1112(5) & 1.86536(5) \\
 & 0.0337 & 0.2941(1) & 0.85 & 1.3190(5) & 0.1123(6) & 2.14981(5) \\
\hline
4 & 0.0228 & 0.2062(2) & 0.273 & 0.5935(2) & 0.0750(3) & 0.8994(1) \\
 & 0.0228 & 0.2062(2) & 0.564 & 0.9313(5) & 0.0754(6) & 1.52542(6) \\
 & 0.0228 & 0.2062(2) & 0.705 & 1.0811(8) & 0.0747(8) & 1.80845(6) \\
 & 0.0228 & 0.2062(2) & 0.85 & 1.228(1) & 0.074(1) & 2.08753(6) \\
\hline
5 & 0.0165 & 0.1548(1) & 0.195 & 0.4427(3) & 0.0555(3) & 0.67113(6) \\
 & 0.0165 & 0.1548(1) & 0.5 & 0.8038(8) & 0.055(1) & 1.34477(8) \\
 & 0.0165 & 0.1548(1) & 0.7 & 1.017(1) & 0.053(2) & 1.75189(7) \\
 & 0.0165 & 0.1548(1) & 0.85 & 1.168(2) & 0.052(2) & 2.04296(7) \\
\end{tabular}\end{ruledtabular}
\end{table}

In our simulations for this paper, we computed decay constants and 
masses for non-physical $H_s$ mesons composed of an $s$~quark, and 
heavy quarks~$h$ with various masses~$m_h$ ranging from below the 
$c$~mass to just below the $b$~mass. This data allows us to extrapolate 
to the $b$~mass, where $m_{H_s}\!=\!m_{B_s}$. We repeated our analysis 
for five different lattice spacings, allowing us also to extrapolate 
our results to zero lattice spacing.

\begin{figure*}
\begin{center}
\includegraphics[scale=0.9]{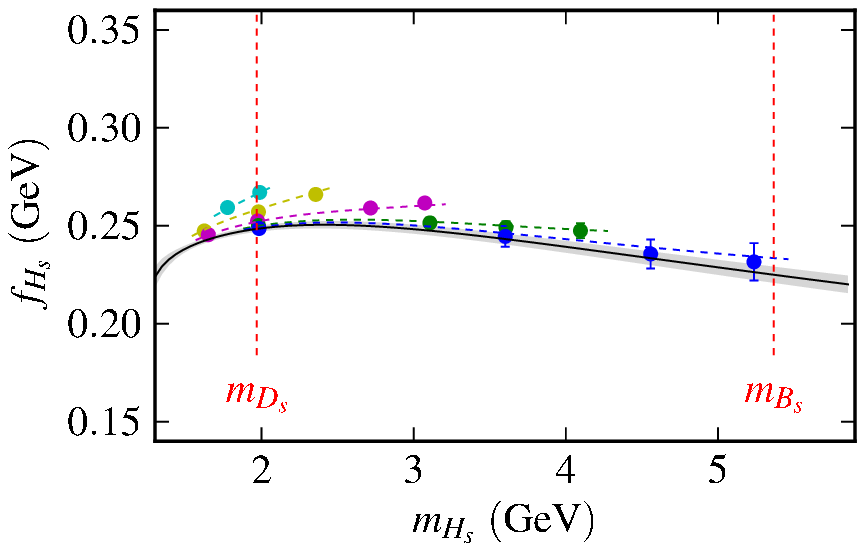}%
\includegraphics[scale=0.9]{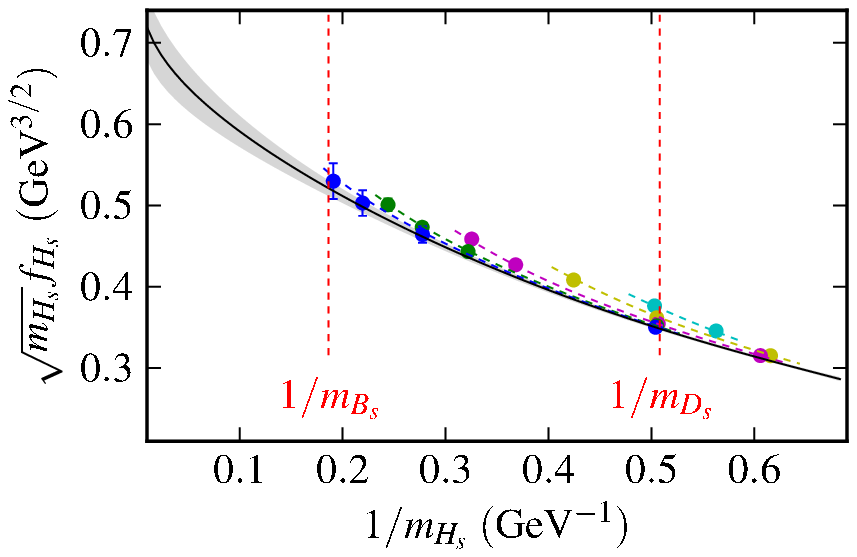}
\end{center}
\caption{\label{fig-fHs}
The leptonic decay constant $f_{H_s}$ for pseudoscalar~$h\overline{s}$ 
mesons~$H_s$, plotted on the left versus the $H_s$~mass as the 
$h$-quark's mass is varied. The solid line and gray band show our 
best-fit estimates for the decay constants extrapolated to zero lattice 
spacing. Best-fit results (dashed lines) and simulation data are also 
shown for five different lattice spacings, with results for smaller 
lattice spacings extending to higher masses (since we restrict 
$am_h\!<\!1$). The simulation data points have been corrected for small 
mistunings of the $s$~quark's mass. On the right the same simulation 
data and fits are plotted for~$\sqrt{m_{H_s}}f_{H_s}$ 
versus~$1/m_{H_s}$.
}
\end{figure*}

The gluon configuration sets we used are from the MILC 
collaboration~\cite{MILC} and are described in Table~\ref{tab-cfg}. Our 
simulation results for the decay constants and meson masses, for 
various values of the $h$~mass, are presented in Table~\ref{tab-data}. 
We also give results for the mass of the pseudoscalar $h\overline{h}$ 
meson, $m_{\eta_h}$, and for the mass of the $\eta_s$ meson. The 
$\eta_s$ is an unphysical pseudoscalar $s\overline{s}$ meson whose 
valence quarks are not allowed to annihilate; we use its mass to tune 
the bare mass of the $s$~quark: simulations show that its mass is 
$m_{\eta_s,\mathrm{phys}}=0.6858(40)$\,GeV when the $s$ mass is 
correctly tuned~\cite{Davies:2009tsa}.

We expect some statistical correlation between results from the same 
configuration set but with different $h$-quark masses. We have not 
measured these, but we have verified that our results are insensitive 
(at the level of $\pm\sigma/4$) to such correlations.
We introduce a 50\%~correlation for our fits, which increases our final 
error estimates slightly.

Our strategy for extracting $f_{B_s}$ is first to fit our simulation 
results for $f_{H_s}$ to the HQET-inspired formula~\cite{bib-hqet}
\begin{align}
   f_{H_s}(a,&\,m_{H_s},m_{\eta_s})
   =\,
   \nonumber\\
   & (m_{H_s})^b    
\left(\frac{\alpha_V(m_{H_s})}{\alpha_V(m_{D_s})}\right)^{-2/\beta_0}
   \sum_{i=0}^{N_m-1} C_i(a)
    \left(\frac{1}{m_{H_s}}\right)^i
    \nonumber \\
   &+c_s (m_{\eta_s}^2-m_{\eta_s,\mathrm{phys}}^2),
    \label{eq-fHs}
\end{align}
where $\beta_0 \!=\! 11-2n_f/3 = 9$ in our 
simulations\,\cite{anom-dim}, $\alpha_V$ is the QCD coupling in the 
$V$~scheme~\cite{Davies:2008sw,McNeile:2010ji}, and constant 
$b\!=\!-0.5$ from HQET. Here we use $m_{H_s}$ as a proxy for the 
$h$-quark mass since its value for $b$~quarks is known from experiment. 
The last term in~\eq{eq-fHs} corrects for tuning errors in the 
$s$-quark mass; we determined that $c_s\!=\!0.06(1)$ by repeating our 
calculations with slightly different $s$~masses~\cite{bib:cs}. We 
parameterize
\begin{equation}\label{eq-cijkl}
   C_i(a) = \sum_{j,k,l=0}^{N_a-1}c_{ijkl}
   \left(\frac{am_h}{\pi}\right)^{2j}
   \left(\frac{am_s}{\pi}\right)^{2k}
   \left(\frac{a\Lambda_\mathrm{QCD}}{\pi} \right)^{2l}
\end{equation}
with $N_m\!=\!N_a\!=\!4$. This expansion is in powers of quark masses 
and the QCD scale parameter~$\Lambda_\mathrm{QCD}\!\approx\!0.5$\,GeV 
divided by the ultraviolet cutoff for the lattice theory: 
$\Lambda_\mathrm{UV}\!\approx\!\pi/a$. The fit parameters are the 
coefficients $c_{ijkl}$ for each of which we use a prior of~$0\pm1.5$, 
which is conservative~\cite{bib-empbayes}.

Our data for five different lattice spacings and a wide range of 
masses~$m_{H_s}$ are presented with our fit results in 
Fig~\ref{fig-fHs}.
The reach in~$m_{H_s}$ grows as the lattice spacing
decreases (since we restrict $am_h<1$), and deviations
from the continuum curve get smaller. The fit is excellent, with a 
$\chi^2$~per degree of freedom of~$0.36$ while fitting all 
17~measurements. The small~$\chi^2$ results from our conservative 
priors (we get excellent fits and smaller errors  with priors that are 
half the width).

Having determined the parameters in \eq{eq-fHs}, the second step in our 
analysis is to set $M_{H_s}\!=\!M_{B_s}$, $a\!=\!0$, and 
$m_{\eta_s}\!=\!m_{\eta_s,\mathrm{phys}}$ in that formula to obtain our 
final value for $f_{B_s}$,
\begin{equation}
   f_{B_s} = 0.225(4)\,\mathrm{GeV},
\end{equation}
which agrees well with the previous best NRQCD result 
of~0.231(15)\,GeV~\cite{Gamiz:2009ku} but is almost four times more 
accurate. Our result also agrees with the recent result 
of~0.232(10)\,GeV from the ETM~collaboration, although that analysis 
includes only two of the three light quarks in the quark 
sea~\cite{etm:2011gx}.

The error budget for our result is given in Table~\ref{tab-errors} and 
shows that the dominant errors come from statistical uncertainties in 
the simulations, the $m_{H_s}\!\to\!m_{B_s}$ extrapolation, the 
$a^2\!\to\!0$ extrapolation, and uncertainties in the scale-setting 
parameter~$r_1$. Our analysis of $f_{D_s}$ in~\cite{Davies:2010ip} 
indicates that finite volume errors, errors due to mistuned sea-quark 
masses, errors from the lack of electromagnetic corrections, and errors 
due to lack of $c$~quarks in the sea are all significantly less 
than~1\%, and so negligible compared with our other uncertainties. Our 
final result is also insensitive to the detailed form of the fit 
function; for example, doubling the number of terms has negligible 
effect ($0.03\sigma$) on the errors and value.

\begin{table}\begin{ruledtabular}
   \caption{\label{tab-errors}
   Dominant sources of uncertainty in our determinations of the $B_s$ 
decay
   constant and the $B_s-\eta_b$~mass difference. Contributions are
   shown from the extrapolations in $m_{H_s}$, $a^2$ and $m_s$, as well
   as statistical errors in the simulation data and errors associated 
with
   the scale-setting parameter $r_1$. Other errors are negligible.}
   \begin{tabular}{rccc}
      & $f_{B_s}$ & $m_{B_s}-m_{\eta_b}/2$\\ \hline
   Monte Carlo statistics
                        &   1.30\%       & 1.49\% \\
   $m_{H_s}\to m_{B_s}$ extrapolation
                        &   0.81     & 0.05 \\
      $r_1$ uncertainty
                        &   0.74       & 0.33 \\
   $a^2\to0$ extrapolation
                        &   0.63       & 0.76 \\
   $m_{\eta_s}\to m_{\eta_s,\mathrm{phys}}$  extrapolation
                        &   0.13       & 0.18 \\
      $r_1/a$ uncertainties
                        &   0.12       & 0.17 \\
      \hline
     Total              &   1.82\%     & 1.73\%
\end{tabular}
\end{ruledtabular}\end{table}

We have also included in Fig.~\ref{fig-fHs} (right) a plot of 
$\sqrt{m_{H_s}}f_{H_s}$ for different values of~$m_{H_s}$. This shows 
that there are large non-leading terms in $f_{H_s}$, beyond the leading 
$1/\sqrt{m_{H_s}}$ behavior predicted by~HQET. Our simulation 
nevertheless provides evidence for the leading term. Treating exponent 
$b$ in \eq{eq-fHs} as a fit parameter, rather than setting it equal 
to~$-0.5$, we find a best-fit value of $b\!=\!-0.51(13)$, in excellent 
agreement with the HQET~prediction. This is the first empirical 
evidence for this behavior.

Our analysis also yields a value for~$f_{D_s}$, which agrees 
with~\cite{Davies:2010ip}. It is also clear from Fig.~\ref{fig-fHs} 
(left)
that $f_{H_s}$ peaks between $f_{D_s}$ and $f_{B_s}$, and that 
$f_{B_s}$ is smaller\,---\,we find:
\begin{equation}
   f_{B_s}/f_{D_s}\!=\!0.906(14).
\end{equation}
HQET suggests a ratio less than one, but previous lattice QCD results 
have been ambiguous about this point.

To check our $f_{Bs}$ analysis technique, we adapted the same technique 
to compute the mass difference~\cite{Gregory:2010gm}
\begin{equation}
   \Delta \equiv m_{H_s} - m_{\eta_h}/2,
\end{equation}
using, as inputs, the masses $m_{\eta_h}$ computed in our simulations 
for pseudoscalar $h\overline{h}$ mesons made of our heavy quark. Our 
values for $\Delta$ come from the results in~Table~\ref{tab-data}. We 
fit them to
\begin{align}
   \Delta(a,&\,m_{H_s},m_{\eta_s}) =
   \nonumber \\
  & m_{H_s}\sum_{i=0}^{N_m-1}D_i(a)
   \left(\frac{1}{m_{H_s}}\right)^i
   +d_s(m_{\eta_s}^2-m_{\eta_s,\mathrm{phys}}^2)
   \label{eq-mHs}
\end{align}
where our simulations indicate that $d_s=0.18(1)$~\cite{bib:ds}, and 
$D_i(a)$ has an expansion similar to that for $C_i(a)$ (\eq{eq-cijkl}), 
with the same priors.

\begin{figure}
\begin{center}
   \includegraphics[scale=0.9]{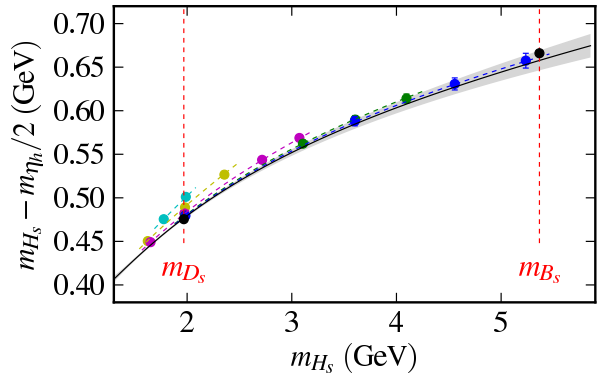}
\end{center}
\caption{\label{fig-mHs}
The $H_s-\eta_h/2$~mass difference plotted versus $m_{H_s}$ as the 
$h$~quark's mass is varied. The solid line and gray band show our 
best-fit estimates for the mass differences extrapolated to zero 
lattice spacing. Best-fit results (dashed lines) and simulation data 
are also shown for five different lattice spacings, with results for 
smaller lattice spacings extending to larger masses (since we require 
$am_h<1$). The simulation data points have been corrected for small 
mistunings in the $s$~quark's mass. Data points (in black) at $m_{D_s}$ 
and $m_{B_s}$ are the experimental values after correcting for small 
effects from electromagnetism, $\eta_b$~annihilation, and 
$c$~quarks in the sea, none of which are included in the simulation.
}
\end{figure}

We show the best fit to our simulation data in~Fig.~\ref{fig-mHs}. 
Again, the fit is excellent, which a $\chi^2$ per degree of freedom 
of~0.13 while fitting all 17~measurements. Extrapolating to the 
$b$~mass, we obtain our best fit value for the mass splitting,
\begin{equation} \label{eq-mbs}
   m_{B_s} - m_{\eta_b}/2 = 0.658(11)\,\mathrm{GeV},
\end{equation}
which agrees well with experiment: experiment gives 0.671(2)\,GeV which 
becomes 0.666(4) after removing corrections from electromagnetism, 
$\eta_b$~annihilation, and $c$~quarks in the sea (not included in our 
simulations)~\cite{bib-pdg,Gregory:2010gm}. Our fit also gives a value 
for $m_{D_s}-m_{\eta_c}/2$; it also agrees well with 
experiment~\cite{Davies:2010ip}.

In this paper, we have shown how to use a highly improved 
discretization of the relativistic quark action to make accurate 
calculations for mesons containing $b$~quarks. Our result for the 
$B_s$~decay constant, $f_{B_s}\!=\!0.225(4)$\,GeV, agrees with other 
determinations from lattice QCD but is almost three times more accurate 
than the most precise previous result. The reliability of our 
extrapolations is underscored both by our previous determination of the 
$b$-quark's $\overline{\mathrm{MS}}$~mass, which agrees with other 
determinations to within errors of less than~$\pm1$\%, and by our 
calculation here of the $m_{B_s}-m_{\eta_b}/2$~mass difference, which 
agrees with experiment to within errors of~$\pm11$\,MeV or less 
than~2\%. Our analysis of the decay constant gives the most extensive 
information to date on the heavy-quark mass dependence of the decay 
constant, and provides the first empirical evidence for the leading 
$1/\sqrt{m_h}$~dependence predicted by HQET. Further results on the 
$B_c$ mesons, and the decay constant~$f_{\eta_h}$ will be presented 
elsewhere~\cite{bib:Bcf}.

Our analysis has important implications for future lattice simulations 
of $B$~physics. Other $B_s$ quantities, like semileptonic form factors, 
can be analyzed in the same way, bringing few-percent precision within 
reach. Similar precision for $B$ quantities is possible by combining 
$B_s$ calculations like these with precise calculations of $B_s/B$ 
ratios using (very efficient) non-relativistic effective field theories 
for $b$-quark dynamics.

We are grateful to the MILC collaboration
for the use of their configurations.
We thank Jack Laiho and Junko Shigemitsu for
useful discussions.
This work was funded by the STFC, the Scottish Universities Physics
Alliance, the NSF (grant PHY-0757868), the MICINN (under grants 
FPA2009-09638 and FPA2008-10732),
the DOE (grant DE-FG02-04ER41299), the DGIID-DGA (grant 2007-E24/2) and 
by the EU under
ITN-STRONGnet (PITN-GA-2009-239353). EF is supported on the MICINN 
Ram\'on
y Cajal program.
Computing time for this project came from
the Argonne leadership Computing Facility at the
Argonne National Laboratory supported by
the Office of Science at the US DoE under
Contract nos. DOE-AC02-06CH11357,
the DEISA Extreme Computing Initiative co-funded through
EU FP6 project RI-031513 and FP7 project
RI-222919, NERSC and the Ohio Supercomputing Centre.
We used chroma for part of our analysis~\cite{bib:chroma}.

\bibliographystyle{plain}

\begin{thebibliography}{16}
\bibitem{anom-dim} We include only the leading-order anomalous 
dimension because of the limited precision of our data. Our results are 
essentially unchanged if we omit leading-order as well.
\end{thebibliography}

\end{document}